\documentclass[fleqn,twoside]{article}
\usepackage{espcrc2,psfig}

\newcommand{\AmS}{{\protect\the\textfont2
  A\kern-.1667em\lower.5ex\hbox{M}\kern-.125emS}}

\title{Spectral density calculations in a heavy-light meson-meson system
\thanks{This material is based upon work supported by the
National Science Foundation under Grant No. PHY-0073362.}}

\author{H. R. Fiebig\address{Physics Department, 
                            FIU-University Park, 
                            Miami, Florida 33199, USA} (LHPCollaboration)}
       
\begin{document}

\begin{abstract}
A system of two static quarks, at
fixed distances r, and two light quarks is
studied on an anisotropic lattice. Excitations
by operators emphasizing quark or gluon degrees
of freedom are examined. The maximum entropy
method is applied in the spectral analysis. These
simulations ultimately aim at learning about
mechanisms of hadronic interaction.
\end{abstract}

\maketitle

\section{INTRODUCTION}

Features of hadronic interaction can in principle be gleaned from
two-hadron energy spectra in finite boxes \cite{Lue91a}.
For two hadrons containing one heavy (static) quark each
\cite{Mih97,Mic99}, the
relative distance $r$ is well defined, and the excitation spectra
as functions of $r$ should give insight into the physics of the strong
interaction.

Refined analysis techniques able to deal with excitations are desirable.
We have employed Bayesian inference \cite{Jar96}
to extract energy spectra from
correlation functions of a set of meson-meson operators.
In lattice QCD Bayesian techniques had not been used until only
very recently \cite{Nak00}. Experience with this analysis
tool is thus somewhat limited. In this contribution we present
selected results
from Bayesian curve fitting with an entropic prior.

\section{OPERATORS AND LATTICE}

The following operators describe systems of two pseudo scalar mesons
in the $I=2$ channel
\begin{eqnarray}
\lefteqn{\Phi_1(t)=\sum_{\vec{x},\vec{y}}
\delta_{\vec{r},\vec{x}-\vec{y}}} \label{Phi1}\\
& &\overline{Q}_{A}(\vec{x}t) \gamma_5 q_{A}(\vec{x}t)\,
\overline{Q}_{B}(\vec{y}t) \gamma_5 q_{B}(\vec{y}t) \nonumber\\
\lefteqn{\Phi_2(t)=\sum_{\vec{x},\vec{y}}
\delta_{\vec{r},\vec{x}-\vec{y}}} \label{Phi2}\\
&& U_{P; AA^\prime}(\vec{x}t,\vec{y}t)\,
U^\dagger_{P; B^\prime B}(\vec{x}t,\vec{y}t) \nonumber\\
&& \overline{Q}_{ A}(\vec{x}t) \gamma_5 q_{ B}(\vec{x}t)\,
\overline{Q}_{ B^\prime}(\vec{y}t) \gamma_5 q_{ A^\prime}(\vec{y}t)\,.
\nonumber
\end{eqnarray}
Heavy and light quark fields are denoted by $Q$ and $q$, respectively,
and $A,A^\prime,B,B^\prime$ are color indices.

Meson fields in $\Phi_1$ 
are local. Those in $\Phi_2$ are spread out over a distance $\vec{r}$
connected by link variable products $U_{P}$ along straight
spatial paths $P$ between $\vec{x}$ and $\vec{y}$, within time slice $t$.

Among the building blocks of the $2\times 2$ correlation matrix
$C_{ij}(t,t_0) =
\langle\hat{\Phi}_i^\dagger(t)\hat{\Phi}_j(t_0)\rangle$
are light-quark propagators, which are computed from random $Z_2$-source
estimators living on time slice $t_0$ only, and heavy-quark propagators,
which are treated in the static approximation.
Gaussian smearing of the quark fields and APE fuzzing of the gauge field
links is applied.

Simulations are done on a $L^3\times T=10^3\times 30$ anisotropic lattice
with a (bare) aspect ratio of $a_s/a_t=3$.
We use the tadpole improved gauge field action of \cite{Mor99} with
$\beta=2.4$. This corresponds to a spatial lattice constant of
$a_s\simeq 0.25{\rm fm}, a_s^{-1}\simeq 800{\rm MeV}$.
The (quenched) anisotropic Wilson fermion action is augmented with
a clover term limited to spatial planes. Only spatial directions are
improved with renormalization factors
$u_s=\langle\, \framebox(5,5)[t]{}\,\rangle^{1/4}$, while $u_t=1$
in the time direction.
The Wilson hopping parameter $\kappa=0.0679$ leads to the mass
ratio $m_\pi/m_\rho \simeq 0.75$.

\section{SPECTRAL DENSITY}
Consider an operator combination $\Phi_v=v_1\Phi_1+v_2\Phi_2$
and the corresponding time correlation function
$C_v(t,t_0)=\langle\hat{\Phi}^\dagger_v(t)\hat{\Phi}_v(t_0)\rangle$.
We will fit an $\omega$-discretization of the spectral model
\begin{equation}
F(\rho_T|t,t_0)=\int_{-\infty}^{+\infty}d\omega\,
\rho_T(\omega)\exp_T(\omega,t-t_0)\,,
\end{equation}
where $\exp_T(\omega,t)=e^{-\omega t+\omega T\Theta(-\omega)}$.
The spectral density $\rho_T(\omega)$ is a discrete sum of $\delta$-peaks
\begin{equation}
\rho_T(\omega)=\sum_{n\neq 0}\delta(\omega-\omega_n)\,
|\langle n|\hat{\Phi}_v(t_0)|0\rangle|^2\,c_n\,,
\end{equation}
here $c_n=e^{-\omega_n T\Theta(-\omega_n)}$.
The $\chi^2$-distance  of the spectral model from the lattice data,
involving $C_v(t_1,t_0)-F(\rho_T|t_1,t_0)$, is computed with the
full covariance matrix between all time slices $t_1\neq t_0$.

We use the discretization
$\Delta\omega=0.04$, $\omega_k=\Delta\omega k$, $k=-50\ldots +125$,
and $\rho_k=\Delta\omega\,\rho_T(\omega_k)$.

Assuming minimal information about $\rho$ the Bayesian prior
probability is $\propto e^{-\alpha S}$, where
$\alpha$ is a parameter and
\begin{equation}
S=\sum_k ( \rho_k-m_k-\rho_k\ln\frac{\rho_k}{m_k})
\end{equation}
is the entropy relative to a default model $m$ \cite{Jar96}.
Minimizing the functional
\begin{equation}
W[\rho\/]=\chi^2/2-\alpha\,S
\end{equation}
then yields the most likely spectral density distribution $\rho$.
Unlike in \cite{Nak00} our implementation is based on 
\begin{equation}
Z_W=\int[d\rho]\,e^{-\beta_W W[\rho]}
\end{equation}
using simulated annealing, or cooling, $\beta_W^{-1}\rightarrow 0$.
Local Metropolis updates $\rho_k\rightarrow x\rho_k$
where $x$ is randomly drawn from the p.d.f. $p_2(x)=x e^{-x}$ are employed.

\section{RESULTS}
\begin{figure} \centering
\psfig{figure=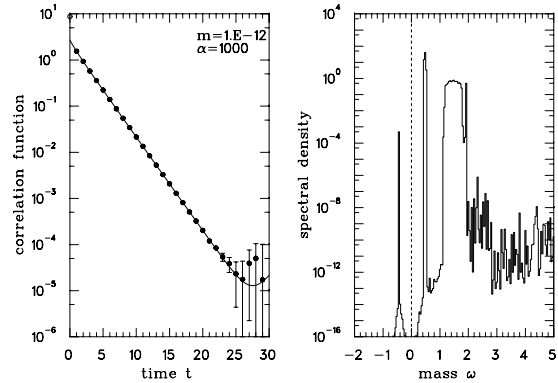,width=72mm,angle=90}
\caption{Single-meson time correlation function and spectral density with a
constant default model $m=10^{-12}$ and entropy parameter $\alpha=1000$.}
\label{fig1} \end{figure}

Analyzing the correlation function of a single local heavy-light meson operator,
see Fig.~\ref{fig1}, reveals two distinct spectral peaks for $\omega>0$.
The dominant narrow peak corresponds to the wide $\log$-linear stretch
of the correlation function. 
The smaller and wider peak originates with data on a few early time slices,
as closer examination shows.
This attests to the sensitivity of the method.
The nature of this excited state is not clear.

\begin{figure}[ht] \centering
 \psfig{figure=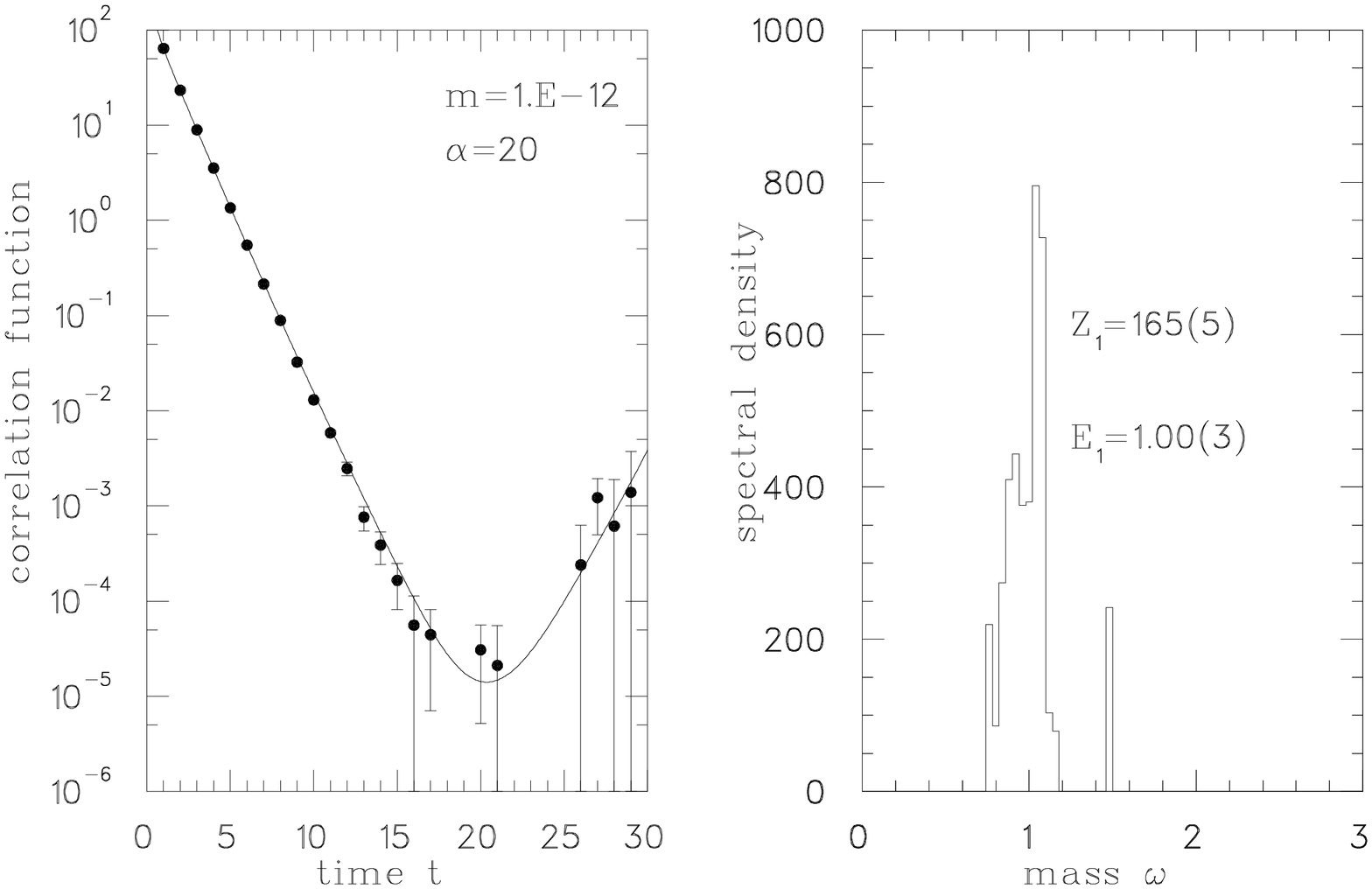,width=72mm,angle=90}\vspace{3.5mm}
 \psfig{figure=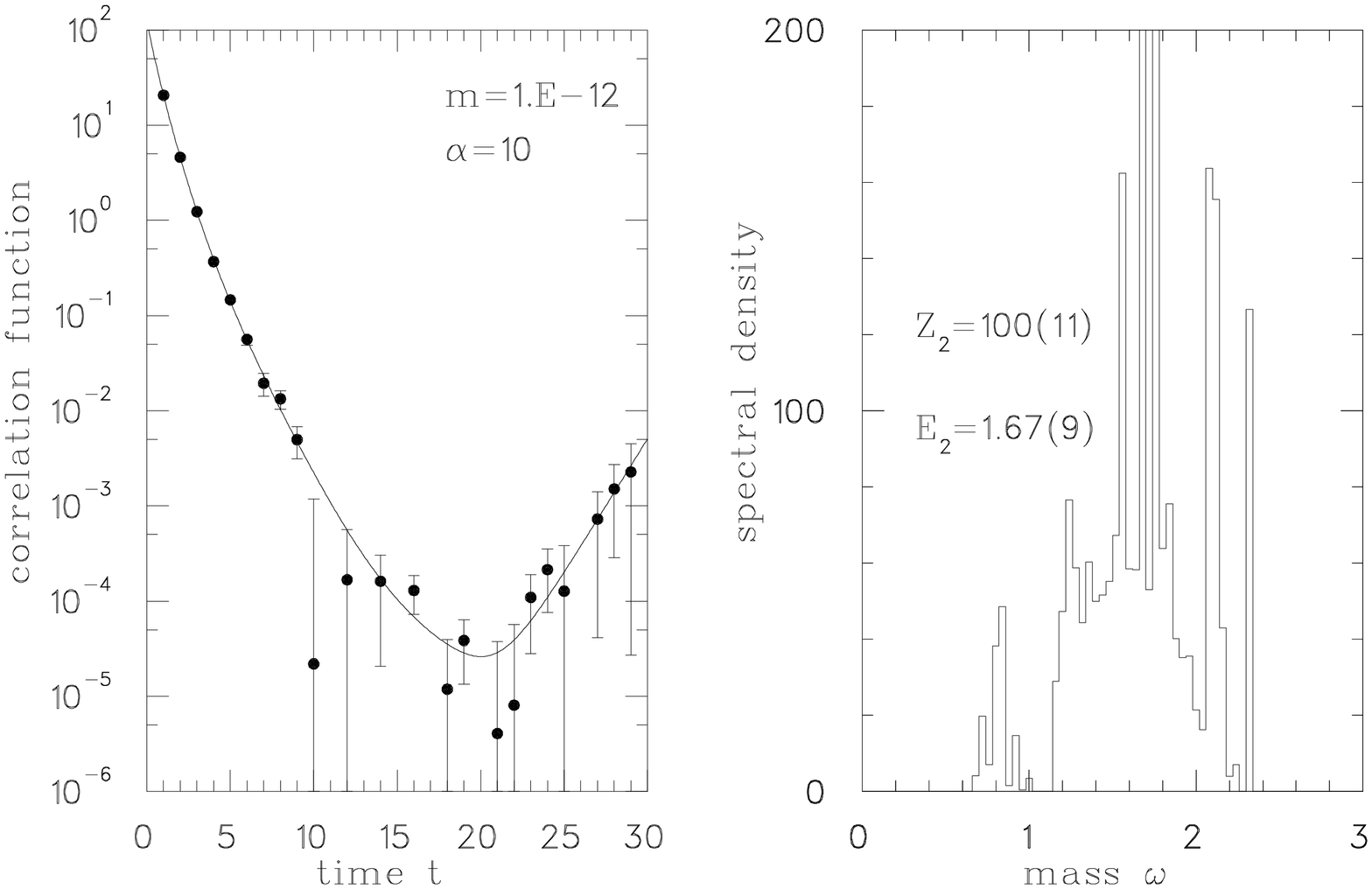,width=72mm,angle=90}\vspace{-3.0mm}
\caption{Two-meson correlators and spectral densities at $r=4$.
The top and bottom pair of figures correspond to $\Phi_1$ and $\Phi_2$,
respectively.}
\label{fig23} \end{figure}

The examples of spectral densities shown in Fig.~\ref{fig23} correspond to
the two-meson operators (\ref{Phi1}) and (\ref{Phi2}), respectively, both at
relative distance $r=4$. 
The nonlocal operator data are noisy to an extent that
precludes the use of standard plateau methods. Bayesian inference still
yields a spectral peak, though broad. 
Empirically, the length of a $\log$-linear stretch (plateau) and the
width of the corresponding peak appear to be inversely related.
The spectral densities in Fig.~\ref{fig23} stem from
averaging over four annealing starts.
The spikes tend to smooth out for a large number of starts. 

There are three main issues with the current approach:
Dependence of the results on the entropy weight parameter $\alpha$,
on the default model $m$, and on the start configuration for annealing.

The minimum of the functional $W[\rho\/]$ is unique.
Different annealing start configurations $\rho$ 
will however move only into the vicinity of the minimum. 
(Information about the shallowness
of $W$ at the minimum may thus be inferred.)
It turns out that integrated quantities like the peak volume and the peak
energy
\begin{eqnarray}
Z_n&=&\sum_{k\in\Delta_n} \rho_k = |\langle n|\hat{\Phi}_v(t_0)|0\rangle|^2\label{Zn}\\
E_n&=&Z_n^{-1}\sum_{k\in\Delta_n} \rho_k\,\omega_k\,, \label{En}
\end{eqnarray}
where $\Delta_n$ is the domain of peak $n$,
are insensitive to annealing starts, whereas the micro structure (fringes)
of the peaks can vary considerably. In this context isolated spikes,
like in Fig.~\ref{fig23}, are common with fine discretization,
but have almost no effect on $Z_n$ and $E_n$.
Those quantities are also extremely stable with respect to changes of $m$ and
$\alpha$. Typical ranges of stability are a remarkable
$m \simeq 10^{-12}\ldots\,10^0$ and $\alpha \simeq 10^{-5}\ldots\,10^{+1}$.

\begin{figure} \centering
\mbox{
\psfig{figure=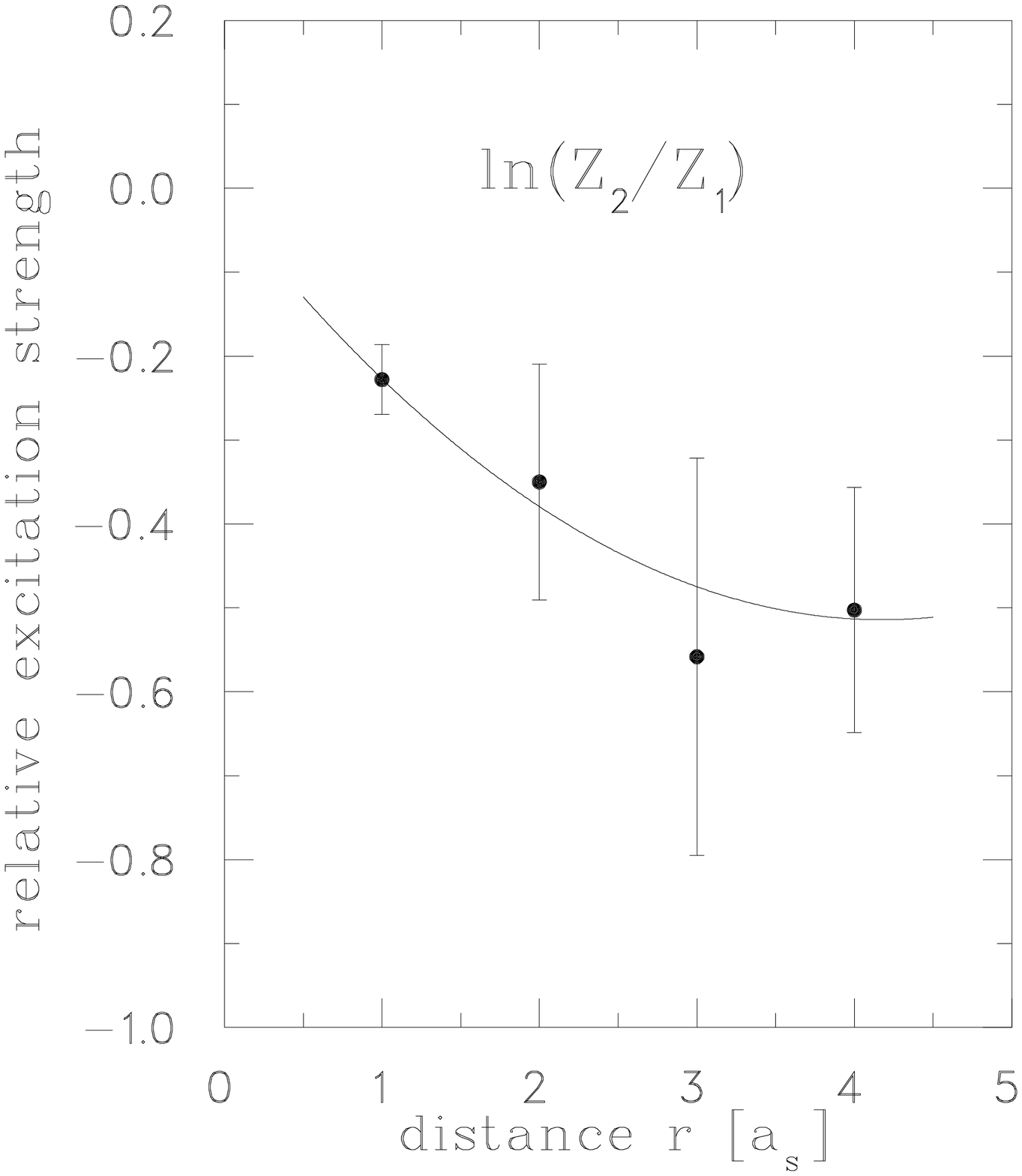,width=35.0mm,angle=0}\rule{2mm}{0mm}
\psfig{figure=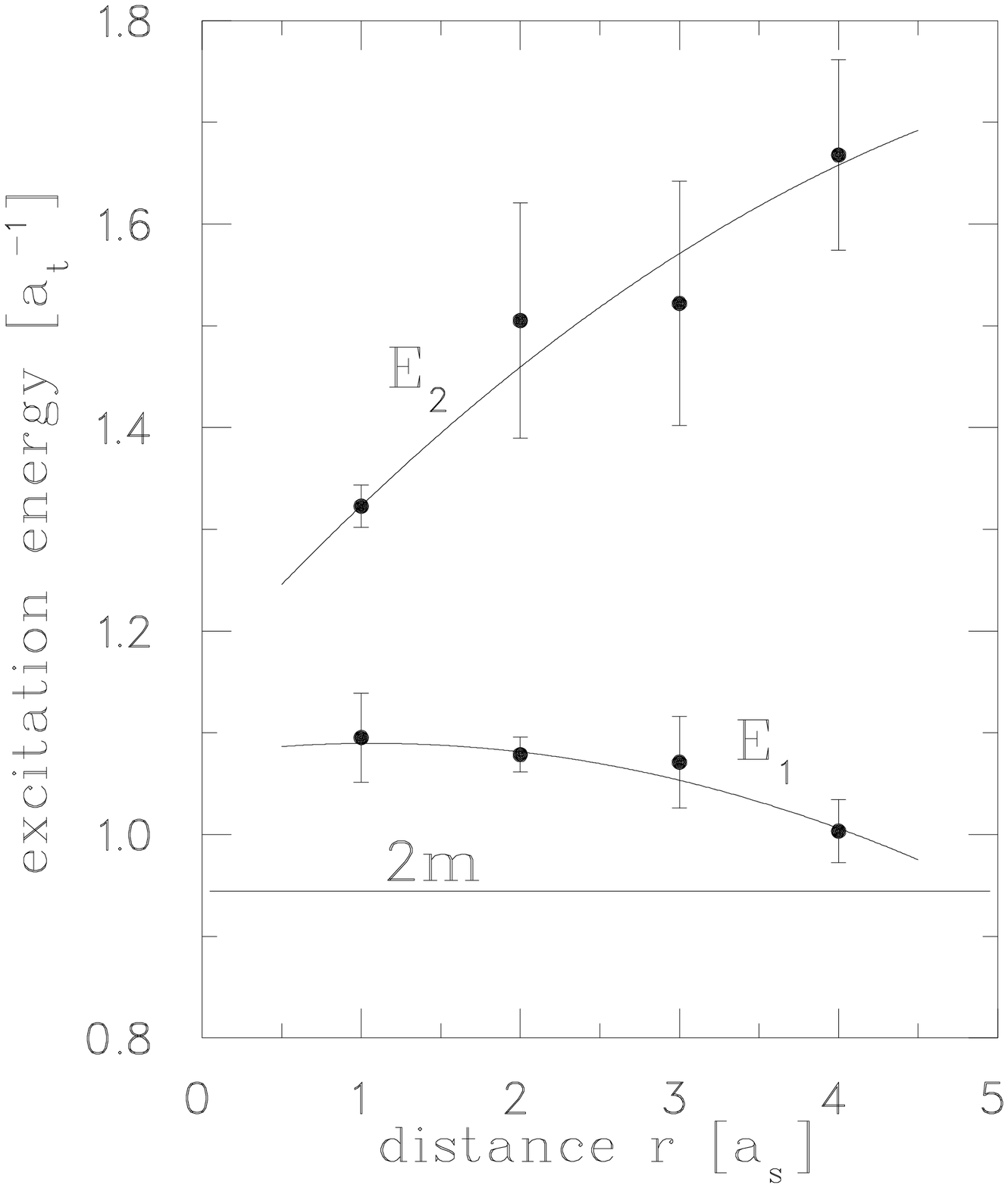,width=34.5mm,angle=0} }\vspace{-3.0mm}
\caption{Comparison of the $\Phi_2$ versus $\Phi_1$ relative excitation strengths
$\ln(Z_2/Z_1)$ and spectral energies $E_{1,2}$ for the ground and excited
states as functions of the relative distance $r$.}
\label{fig4} \end{figure}

We have used 628 gauge configurations. The corresponding statistical
errors on $Z_n,E_n$ are comparable to
fluctuations due to different annealing starts.
Errors bars in Fig.~\ref{fig4} reflect annealing start variances.

Preliminary results for the $r$ dependence of selected quantities are shown
in Fig.~\ref{fig4}. The $Z_2/Z_1$ ratio points at an enhancement of
$\Phi_2$-excitations
(gluon degrees of freedom) over $\Phi_1$-excitations (quark d.o.f.)
as the relative distance is decreased. Those two mechanisms of interaction
are repulsive (in $I=2$), but move in opposite directions with $r$.
They possibly compete at small $r<1$ in the chiral limit \cite{Fie00a}.
The adiabatic potential $V_{\rm ad}(r)={\rm Min}(E_1,E_2)$ is repulsive
for the present set of lattice parameters.

\section{CONCLUSION}

Spectroscopic analysis by way of Bayesian inference
is a very powerful method to treat excitations of hadronic systems.
The excitations of a heavy-light meson-meson give insight into
aspects of strong interaction physics.

\end{document}